\documentclass[twocolumn,showpacs,showkeys,preprintnumbers,amsmath,amssymb,pre,floatfix]{revtex4}

\usepackage{graphicx}% Include figure files
\usepackage{dcolumn}% Align table columns on decimal point
\usepackage{bm}% bold math
\usepackage{psfrag}

\begin{document}

\title{Geometric phases and quantum phase transitions in open systems}

\author{Alexander I. Nesterov}
   \email{nesterov@cencar.udg.mx}
\affiliation{Departamento de F{\'\i}sica, CUCEI, Universidad de Guadalajara,
Av. Revoluci\'on 1500, Guadalajara, CP 44420, Jalisco, M\'exico}

\author{S. G. Ovchinnikov}%
 \email{sgo@iph.krasn.ru}
\affiliation{L. V. Kirensky Institute of Physics, SB RAS, 660036, Krasnoyarsk, Russia and Siberian Federal University, 660041, Krasnoyarsk, Russia}

\date{\today}

\begin{abstract}
The relationship between quantum phase transition and complex
geometric phase for open quantum system governed by the non-Hermitian
effective Hamiltonian with the accidental crossing of the eigenvalues is established. In particular, the geometric phase associated with the ground state of the one-dimensional dissipative Ising model in a transverse magnetic field is evaluated, and it is demonstrated that related quantum phase transition is of the first order.

\end{abstract}



\maketitle

Quantum phase transition (QPT) is characterized by qualitative changes of the ground state of many body system and occur at the zero temperature. QPT being purely quantum phenomena driven by quantum fluctuations is associated with the energy level crossing and implies the lost of analyticity in the energy spectrum at the critical points \cite{SS}. A first order QPT is determined by a discontinuity in the first derivative of the ground state energy. A second order QPT means that the first derivative is continuous, while the second derivative has either a finite discontinuity or divergence at the critical point. Since QPT is accomplished by changing some parameter in the Hamiltonian of the system, but not the temperature, its description in the standard framework of the Landau-Ginzburg theory of phase transitions failed, and identification of an order parameter is still an open problem \cite{SGC}. In this connection, an issue of a great interest is recently established  relationship between geometric phases and quantum phase transitions \cite{PC,CP,Zhu,HA}. This relation is expected since the geometric phase associated with the energy levels crossings has a peculiar behavior near the degeneracy point. It is supposed that the geometric phase, being  a measure of the curvature of the Hilbert space, is able to capture drastic changes in the properties of the ground states in presence of QPT \cite{CP,Zhu,HA,ZSL}.

In this Rapid Communication we analyze relation between the geometric phase and QPT in an open quantum system governed by non-Hermitian Hamiltonian. We found that QPT is closely connected with the geometric phase and the latter may be considered as an universal order parameter for description of QPT. Studying the dissipative one-dimensional Ising model in a transverse magnetic field we demonstrated that the QPT being of the second order in absence of dissipation is of the first order QPT for the open system.

{\em Degeneracy points and geometric phase.--} We consider an open quantum mechanical system which together with its environment forms a closed system. The description of the such systems by effective non-Hermitian Hamiltonian is well known beginning with the classical papers by Weisskopf and Wigner on the metastable states \cite{WW,WW1}\footnote{For discussion and recent development  see e.g. \cite{JMR,RI,RI1}}.

For the Hermitian Hamiltonian coalescence of eigenvalues results in different eigenvectors, and related degeneracy referred to as `conical intersection' is known also as `diabolic point'. However, in a quantum mechanical system governed by non-Hermitian Hamiltonian not only merging of eigenvalues of the Hamiltonian but the associated eigenvectors can be occurred as well. The point of coalescing is called an ``exceptional point". At the latter the eigenvectors merge forming a Jordan block (for review and references see e.g. \cite{B,H1}).

In the context of the Berry phase the diabolic point is associated with `fictitious magnetic monopole' as follows. Assume that for adiabatic driving quantum system two energy levels may cross. Then the energy surfaces form the sheets of a double cone, and its apex is called a ``diabolic point'' \cite{BW}. Since for generic Hermitian Hamiltonian the codimension of the diabolic point is three, it can be characterized by three parameters $\mathbf R= (X,Y,Z)$. The eigenstates $|n,\mathbf R \rangle$ give rise to the Berry's connection defined by ${\mathbf A}_n(\mathbf R)= i\langle n,\mathbf R| \nabla_{\mathbf R} |n,\mathbf R \rangle$, and the curvature
$\mathbf B_n = \nabla_{\mathbf R} \times {\mathbf A}_n $ associated
with ${\mathbf A}_n$ is the field strength  of `magnetic' monopole located at
the diabolic point \cite{B0,BD}. The Berry phase $\gamma_n= \oint_{\mathcal C }{\mathbf A}_n \cdot d \mathbf R$ is interpreted as a holonomy associated with the parallel transport along a circuit $\mathcal C$ \cite{SB}. Similar treatment of the non-Hermitian Hamiltonian yields the `fictitious complex monopole' located at the exceptional point \cite{NF2}.

For the first time, the extension of the Berry phase to the non-Hermitian systems has been done by Garrison and Wright as follows \cite{GW}. Let an adjoint pair $\{|\Psi(t)\rangle, \langle\widetilde\Psi(t)|\}$ be a solution of the time dependent Schr\"odinger equation and its adjoint equation ($\hslash =1$)
\begin{align}\label{S1}
i\frac{\partial }{\partial t}|\Psi(t)\rangle = H(\lambda(t))|\Psi(t)\rangle, \\
-i\frac{\partial }{\partial t}\langle\widetilde\Psi(t)| =
\langle\widetilde\Psi(t)|H(\lambda(t))\label{S2},
\end{align}
where $\lambda \in \mathfrak M$, the parameter space being $\mathfrak M$. Let $|\psi_n(\lambda)\rangle$ and $\langle\tilde\psi_n(\lambda)|$ being right and left eigenvectors of the Hamiltonian: $H(\lambda)|\psi_n(\lambda)\rangle = E_n(\lambda)|\psi_n(\lambda)\rangle$, $\langle\tilde\psi_n(\lambda)| H(\lambda) = E_n(\lambda) \langle\tilde\psi_n(\lambda)| $. Now suppose that there exists a time period $T$ for which $\lambda(T)= \lambda(0)$, then a complex geometric phase $\gamma_n$ is given by the integral \cite{GW,B}
\begin{equation}\label{Eq27}
 \gamma_n =\oint_{\mathcal C}A^{(n)}= i\oint_{\mathcal C}\frac{ \langle\tilde\psi_n(\lambda)|\nabla_a|\psi_n(\lambda)\rangle d\lambda^a}{\langle\tilde\psi_n(\lambda)|\psi_n(\lambda)\rangle}
\end{equation}
where the integration is performed over the contour $\mathcal C$ in the parameter space, $a=1, \dots, \dim\mathfrak M$, $A^{(n)}$ being the connection one-form . Further we assume that the instantaneous eigenvectors form the bi-orthonormal basis, $\langle\tilde\psi_m|\psi_n\rangle =\delta_{mn}$ \footnote{This can alter the definition (\ref{Eq27}) up to the topological contribution $\pi n$, $n\in \mathbb Z$ \cite{MKS}.}\footnote{The geometric phase for systems governed by the non-Hermitian Hamiltonian were studied by various authors, for details and Refs. see e.g. \cite{GW,BD,B,B3,GXQ,H,KM}.}.

{\em  Geometric phase and quantum phase transition. --}  Analysis of the relation between QPT and geometric phase we begin with consideration of a two-level system described by generic non-Hermitian Hamiltonian
$H= \lambda_0 {1\hspace{-.125cm}1}+\mathbf R(t)\cdot \boldsymbol \sigma$,
where $ \sigma_i$ are the Pauli matrices, $\mathbf R(t) = (X,Y,Z)$ is slowly varying and $\lambda_0,X,Y,Z \in \mathbb C$. Using the spinless fermionic creation and annihilation operators, which  obeys an anticommutation relations
$\{C,C\}=0,$ $ \{C^{\dag},C^{\dag}\}=0,$ and $\{C,C^{\dag}\}=1$,
one can rewrite the Hamiltonian as
$H= (\lambda_0 -R){1\hspace{-.125cm}1} + 2R C^{\dag}C$,
where $R={(X^2 +Y^2+Z^2)}^{1/2}$. The ground state $|u_{-}\rangle$ is defined as the vacuum state determined by $C |u_{-}\rangle =0$.

The instantaneous eigenvectors are found to be
\begin{align}
&|u_{-}\rangle = \left(\begin{array}{c}
-e^{-i\varphi}\sin\frac{\theta}{2}\\
\cos \frac{\theta}{2} \end{array}\right), \; \langle \widetilde u_{-}|
=\bigg(-e^{i\varphi}\sin\frac{\theta}{2}, \cos\frac{\theta}{2}\bigg)
\nonumber \\
&|u_{+}\rangle = \left(\begin{array}{c}
                  e^{-i\varphi}\cos\frac{\theta}{2}\\
                  \sin\frac{\theta}{2}
                  \end{array}\right),
\langle \widetilde u_{+}| = \bigg(e^{i\varphi}\cos\frac{\theta}{2},
\sin\frac{\theta}{2}\bigg)  \label{r}
\end{align}
where $\theta, \varphi$ are the complex angles of the complex spherical
coordinates, and the complex energy spectrum is given by $E_{\pm}= \lambda_0 \pm R$. Coupling of eigenvalues occurs when $R =0$ and there are two cases. The first one is of the diabolic point located at the origin coordinates. The second case yields the exceptional point $(X_0,Y_0,Z_0)$. At the latter the eigenvectors coincide up to the phase factor, $|u_{+}\rangle= e^{i\kappa}|u_{-}\rangle$ and $\langle \tilde u_{+}|=e^{-i\kappa}\langle \tilde u_{-}|$ \cite{H1,GRS}.

The geometric phase of the ground state is given by $\gamma = (1/2)\oint_{\mathcal C} q(1 -\cos\theta)d\varphi$, where integration is performed over the contour $\mathcal C$ on the complex sphere $S^2_c$. Let us assume that the contour $\mathcal C$ of integration  is chosen as $\theta= \rm const$. Then the geometric phase
of the ground state is given by
$\gamma = \pi(1- {Z}/{R})$ and can be written as $ \gamma= \pi(1 + \partial E_{-}/\partial Z)$,
where $E_{-}$ is the ground state energy. As can be observed, lost of analyticity occurs at the degeneracy `point' defined by $R=0$ and on the Dirac string attached to the complex fictitious monopole and crossing the complex sphere $S^2_c$ at the south pole.

Further simplification can be made writing $\mathbf R = \mbox{\boldmath$\rho$} - i \mbox{\boldmath$\varepsilon$}$, where we set $\mbox{\boldmath$\rho$}=(x,y,z)$. Without loss of generality we may choose the coordinate system thus, that $\mbox{\boldmath$\varepsilon$}= (0,0,\varepsilon)$. Then computation of
geometric phase yields
\begin{equation}\label{Q3}
\gamma= \pi\bigg(1 - \frac{z -i\varepsilon }{\sqrt{r^2 + (z - i\varepsilon)^2}}\bigg)
\end{equation}
where $r=\sqrt{x^2 + y^2}$.

In what follows we consider the behavior of the geometric phase near the critical points, starting with the Hermitian Hamiltonian. Inserting $\varepsilon=0$ in Eq. (\ref{Q3}), we obtain $\gamma= \pi\big(1 - {z}/(r^2 + z^2)^{1/2}\big)$. This implies that the geometric phase behaves as the step-function near the diabolic point. Considering the general case, we obtain
\begin{align}
 \rm Re \gamma= \Bigg \{
\begin{array}{l}
 \pi, \;{\rm if}\; r > \varepsilon, \, (z=0) \\
  \pi \Big(1\mp \displaystyle\frac{\varepsilon}{\sqrt{\varepsilon^2 - r^2 }}\Big), \; {\rm if} \;
  r < \varepsilon, \, z \rightarrow \pm 0
\end{array}
\end{align}
where the upper/lower sign corresponds to $z \rightarrow \pm 0$,
\begin{figure}[tbh]
\begin{minipage}[]{9 cm}
\begin{center}
\scalebox{0.225}{\includegraphics{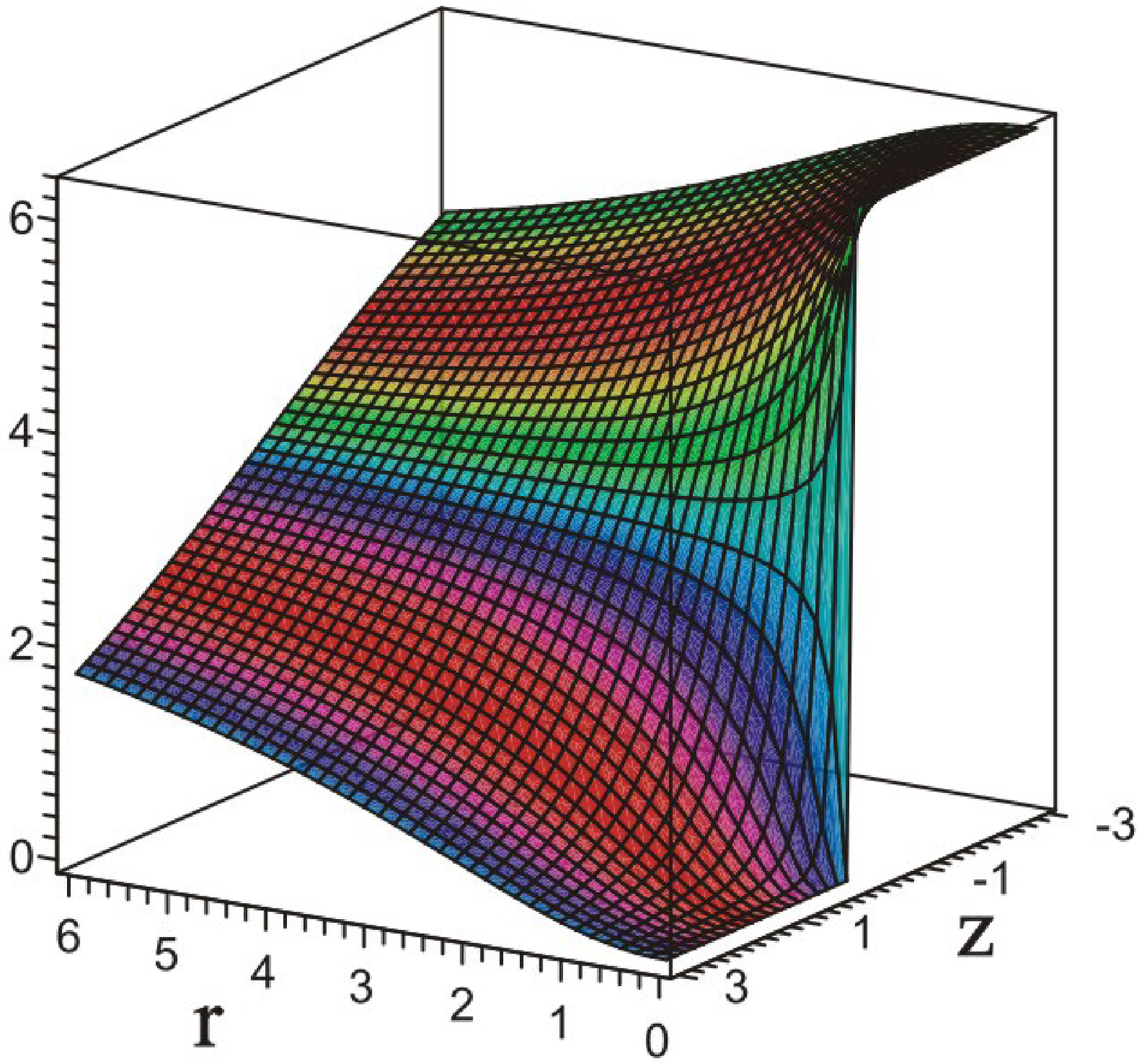}}
%\hspace{-0.75cm}
\scalebox{0.225}{\includegraphics{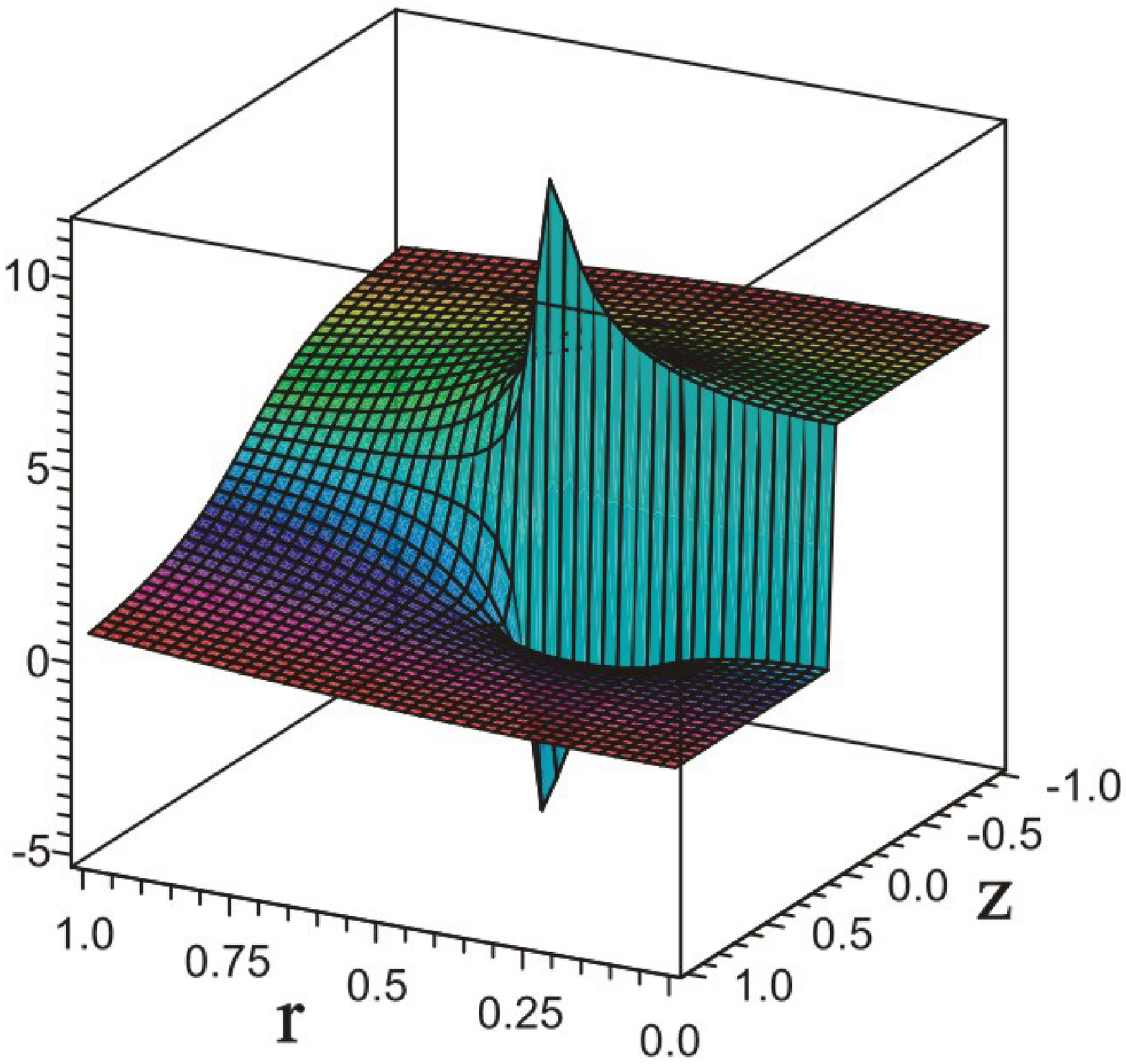}}
\end{center}
\end{minipage}
\caption{Left panel ($\varepsilon =0$): There is clear step function behavior of the geometric phase at the diabolic point $r=z=0$. Right panel: $\rm Re\gamma$ nearby the exceptional point ($\varepsilon =0.5$). }\label{EP1g}
\end{figure}
As can be observed in Fig. \ref{EP1g}, if $\varepsilon =0$ the geometric phase behaves as step-like function near the diabolic point. In addition, $\rm Re \gamma \rightarrow \pm \infty$ at the exceptional point $r=\varepsilon$, and it behaves as a step-like function while $r\rightarrow 0$.
Similar consideration of the imaginary part yields
\begin{align}\label{Q5}
 \rm Im \gamma= \Bigg \{
\begin{array}{l}
 0, \;{\rm if}\; r < \varepsilon \; (z=0)\\
   \displaystyle\frac{\pi\varepsilon}{\sqrt{ r^2 - \varepsilon^2 }}, \; {\rm if} \; r >\varepsilon  \; (z=0)
\end{array}
\end{align}
and clearly it diverges at the exceptional point, $\rm Im \gamma \rightarrow \infty$.

Once return to general non-Hermitian $N$-dimensional problem, consider the
non-Hermitian diagonalizable Hamiltonian $H(\lambda)= \sum_{i=1} E_i |\psi_i\rangle \langle \tilde\psi_i|$. The ground state is given by $|\psi_g (\lambda)\rangle = \otimes^N_{i=1}|\psi_i (\lambda) \rangle$, and computation of geometric phase yields
\begin{align}\label{B0}
\gamma = i\oint_{\cal C} \langle \tilde \psi_g(\lambda)| \frac{\partial}{\partial \lambda^a} |\psi_g(\lambda)\rangle d \lambda^a = \sum^N_{i=1} \gamma_i
\end{align}
where $\gamma_i$ is the geometric phase associated with the eigenvector $|\psi_i(\lambda) \rangle$. Then, applying the Stokes theorem and the Sghr\"odinger equation $H|\chi_m \rangle = E_m |\chi_m\rangle$ together with its adjoint equation $\langle \tilde \chi_m| H = E_m \langle \tilde \chi_m|$, we obtain
\begin{align}
%\label{B0a}
\gamma = -i\sum^N_{i=1} \sum^N_{m\neq i}\iint_{\Sigma}\frac {\langle \tilde \chi_i|\nabla_a H |\chi_m\rangle
 \langle \tilde \chi_m| \nabla_b H |\chi_i\rangle d\lambda^a\wedge d\lambda^b } {(E_m-E_i)^2}
 \nonumber
 \end{align}
It follows herefrom that the curvature $F^{(i)}= dA^{(i)}$ diverges at the degeneracy points, where the energy levels, say $E_n$ and $E_{n+1}$, are crossing, and reaches its maximum values at the avoided level crossing points. Thus, the critical behavior of the system is reflected in the geometry of the Hilbert space through the geometric phase of the ground state.

Since in the neighborhood of either diabolic or exceptional point only terms related to the invariant subspace formed by the two-dimensional Jordan block make substantial contributions, the $N$-dimensional problem becomes effectively two-dimensional (for details see \cite{Arn,KMS}). This implies that there exists the map $\varphi :\mathfrak M \mapsto S^2_c$ such that in the vicinity of the degeneracy points the quantum system can be described by the effective two-dimensional Hamiltonian $H_{ef}= \lambda_0 {1\hspace{-.125cm}1}+\mathbf R\cdot \boldsymbol \sigma$, where $R= (E_{n+1} - E_n)/2$. Then we have
\begin{align}\label{B7}
\gamma \approx \frac{1}{2} \int_{\Sigma^\prime} \frac{\mathbf R \cdot d\mathbf S }{R^3} + \sum_{i\neq n,n+1} \gamma_i(\mathbf R )
\end{align}
where $\Sigma^\prime=\varphi (\Sigma) \subset S^2_c$. The behavior of the geometric phase described by the first term is independent of a peculiarities of quantum-mechanical system. Therefore, one can consider the complex Bloch sphere as an universal parameter space for description of QPT in the vicinity of the critical point.

Following \cite{CP}, we define the overall geometric phase of the ground state as $\gamma_g = (1/N)\sum^N_{i=1} \gamma_i$. In the thermodynamical limit $\gamma_g = \int \gamma(x)d\mu(x)$, where $d\mu(x)$ is the suitable measure.
As has been shown by Zhu \cite{Zhu} on example of $XY$ spin chain, the overall geometric phase associated with the ground state exhibits universality, or scaling behavior in the vicinity of the critical point. In addition, the geometric phase allows to detect the critical point in the parameter space of the Hamiltonian \cite{PC,CP,HA,Zhu,ZSL}. These works indicate that the overall geometric phase $\gamma_g$ can be considered as the universal order parameter for description of QPT.

{\em Geometric phase and QPT in the quantum Ising model.} -- As illustrative example we consider the 1-dimensional Ising model in a transverse magnetic field with dissipation governed by the non-Hermitian Hamiltonian:
\begin{align}
H= - J\sum^N_{n=1}\big(h \sigma^x_n + \sigma^z_n \sigma^z_{n+1} - i\frac{\delta}{2}\sigma^{+}_n \sigma^{-}_n\big)
\end{align}
with the periodic boundary condition $\mbox{\boldmath$\sigma$}_{N+1} =\mbox{\boldmath$\sigma$}_1$. The external magnetic field is described by the parameter $h$ and spontaneous decay is described by $\Gamma= \sqrt{\delta}\sigma^{-}_n$ with a source of decoherence being $ \sigma^{-}_n =  (\sigma^{z}_n -i \sigma^{y}_n)/2$.

To study the geometric phase in this system we consider the more general Hamiltonian $H(h,\delta,\varphi) = g_\varphi Hg^{\dagger}_\varphi$, where $g_\varphi = \prod_{n=1}^N e^{i\frac{\varphi}{4} \sigma_n^x}$ and $0 \leq \varphi < 2\pi$. After applying the standard Jordan-Wigner transformation and following the procedure outlined in \cite{DJ,CP}, we find that the system can be described in terms of the non-interacting quasiparticles with the reduced Hamiltonian
\begin{align}\label{I}
H^{+} =& -J\sum^N_{n=1}\big(c^\dagger_n c_{n+1 }+ e^{i\varphi}c_{n+1} c_{n} + g + i\delta \nonumber \\
& - 2g c^\dagger_n c_{n } + c^\dagger_{n +1}c_{n }+ e^{-i\varphi}c^\dagger_{n} c^\dagger_{n+1} \big)
\end{align}
where $g = h - i \delta$, and $c_n$ are fermionic operators satisfying anticommutation relations $\{c_m,c^\dagger_n\} = \delta_{mn}$ and $\{c_m,c_n\}=\{c^\dagger_m,c^\dagger_n\}=0$. Applying the Fourier transformations $c_n = e^{-i\pi/4}\sum_k c_k e^{ikna}/N^{1/2}$  with the antiperiodic boundary condition $c_{N+1}=-c_1$, we obtain
%\begin{align}\label{Ia}
$H^{+} =J\sum_k\Big(2(g - \cos(ka))c^\dagger_k c_k + \sin(ka)( e^{-i\varphi}c^\dagger_k c^\dagger_{-k} + e^{i\varphi}c_{-k} c_{k} )- g - i \frac{\delta}{2}\Big)$,
%\end{align}
where $k= \pm \pi/Na,\dots, \pm (N-1)\pi/Na$ is a half-integer quasimomentum, the lattice spacing being $a$.

The Hamiltonian $H^{+}$ can be diagonalized by using the Bogoliubov transformation: $c_k = \tilde u_k b_k +  v_{-k}b^\dagger_{-k}$, $
c^\dagger_k =  u_k b^\dagger_k + \tilde v_{-k}b_{-k}$. The Bogoliubov modes $(u_k,v_k)$ and  $(\tilde u_k,\tilde v_k)$ satisfy the Schr\"odinger equation and its adjoint equation, respectively, with the Hamiltonian $H(k) = -{iJ\delta}{1\hspace{-.125cm}1}+ \mathbf R (k)\cdot \boldsymbol \sigma$, and
$\mathbf R(k) = 2J\big(\sin(ka)\cos\varphi,\sin(ka)\sin\varphi , g - \cos(ka)\big)$. There are two eigenstates for each $k$ with the complex energies $\varepsilon_{\pm}(k)= \varepsilon_{0} \pm \varepsilon(k)$, where we set
$\varepsilon_{0} = -iJ\delta $ and $ \varepsilon(k) = 2J{(g^2 -  2g\cos(ka) +1)^{1/2}}$. The positive energy eigenstate
$|u_{+}(k)\rangle = \left(\begin{array}{c}
                 u_k\\
                  v_k
                  \end{array}\right)$, $
\langle \widetilde u_{+}(k)| = \big(\tilde u_k,
\tilde v_k\big)$
normalized so that $\tilde u_k u_k+ \tilde v_k v_k=1$, defines the quasi-particle operators $b_k =  \tilde u_k c_k +  \tilde v_{k}c^\dagger_{-k}$ and $ b^\dagger_k =   u_k c^\dagger_k +  v_{k}c_{-k}$ as follows:
$b_k = e^{i\varphi}\cos\frac{\theta_k}{2} \, c_k +  \sin\frac{\theta_k}{2} \,c^\dagger_{-k}$, $b^\dagger_k = e^{-i\varphi}\cos\frac{\theta_k}{2} \, c^\dagger_k + \sin\frac{\theta_k}{2} \, c_{-k}$, where
\begin{align}\label{Th}
\cos\theta_k = \frac{ g -  \cos(ka)}{\sqrt{g^2 -  2g\cos(ka) +1}}.
\end{align}

Using these results, we obtain the diagonalized Hamiltonian as a sum of quasi-particles with half-integer quasimomenta, $H^{+}=  \sum_k\big(\varepsilon_0 +  \varepsilon(k)(b_k^\dagger b_k- \frac{1}{2})\big)$. Its ground state is given as product of qubit-like states:
\begin{align}
&|\psi_g\rangle =  \bigotimes_{k} \Big(\cos\frac{\theta_k}{2}|0\rangle_k  |0\rangle_{-k}  -  e^{-i\varphi}\sin\frac{\theta_k}{2}|1\rangle_k  |1\rangle_{-k} \Big) \nonumber \\
&\langle \tilde\psi_g |=  \bigotimes_{k} \Big(\cos\frac{\theta_k}{2}\langle 0|_k \langle 0|_{-k}  -  e^{i\varphi}\sin\frac{\theta_k}{2}\langle 1|_k \langle 1|_{-k} \Big)
\nonumber
\end{align}
where $|0\rangle_k $ is the vacuum state of the mode $b_k$, and $|1\rangle_k  $ is the first excited state, $|1\rangle_k =b^\dagger_k |0\rangle_k$. Each single state lies in the two-dimensional Hilbert space spanned by $|0\rangle_k  |0\rangle_{-k}$ and $|1\rangle_k  |1\rangle_{-k}$. For given value of $k$ the state in each of these two-dimensional Hilbert space can be presented as the point on the complex two-dimensional sphere $S^2_c$ with coordinates $(\theta_k,\varphi)$.

For $|g|\gg 1$ the ground state is a paramagnet with all spin oriented along the $x$ axis, and  from Eq. (\ref{Th}) we obtain $\cos\theta_k \rightarrow 1$ while $|g| \rightarrow \infty$. Thus, the north pole of the complex Bloch sphere corresponds to  paramagnetic ground state. On the other hand, when $|g|\ll 1$ there are two degenerate ferromagnetic ground states with the all spins polarized up or down along the $z$ axis. The real part of the complex energy reaches its minimum at the point defined by $\cos\theta_k = -1$, and, hence, the south pole of the complex sphere is related to the  pure ferromagnetic ground state with the broken symmetry when all spins have orientation up or down. However, in the thermodynamical limit the system passing through the critical point ends in a superposition of the up and down states with finite domains of spins separated by kinks \cite{DJ}.

The geometric phase of the ground state is found to be
\begin{align}\label{B2}
\gamma = i \int_0^{2\pi}\langle \tilde \psi_g|\frac{\partial}{\partial \varphi}|\psi_g \rangle \, d\varphi = \sum_{k >0}\pi(1 - \cos\theta_k)
\end{align}
\begin{figure}[tbh]
\begin{minipage}[]{9 cm}
\begin{center}
\scalebox{0.215}{\includegraphics{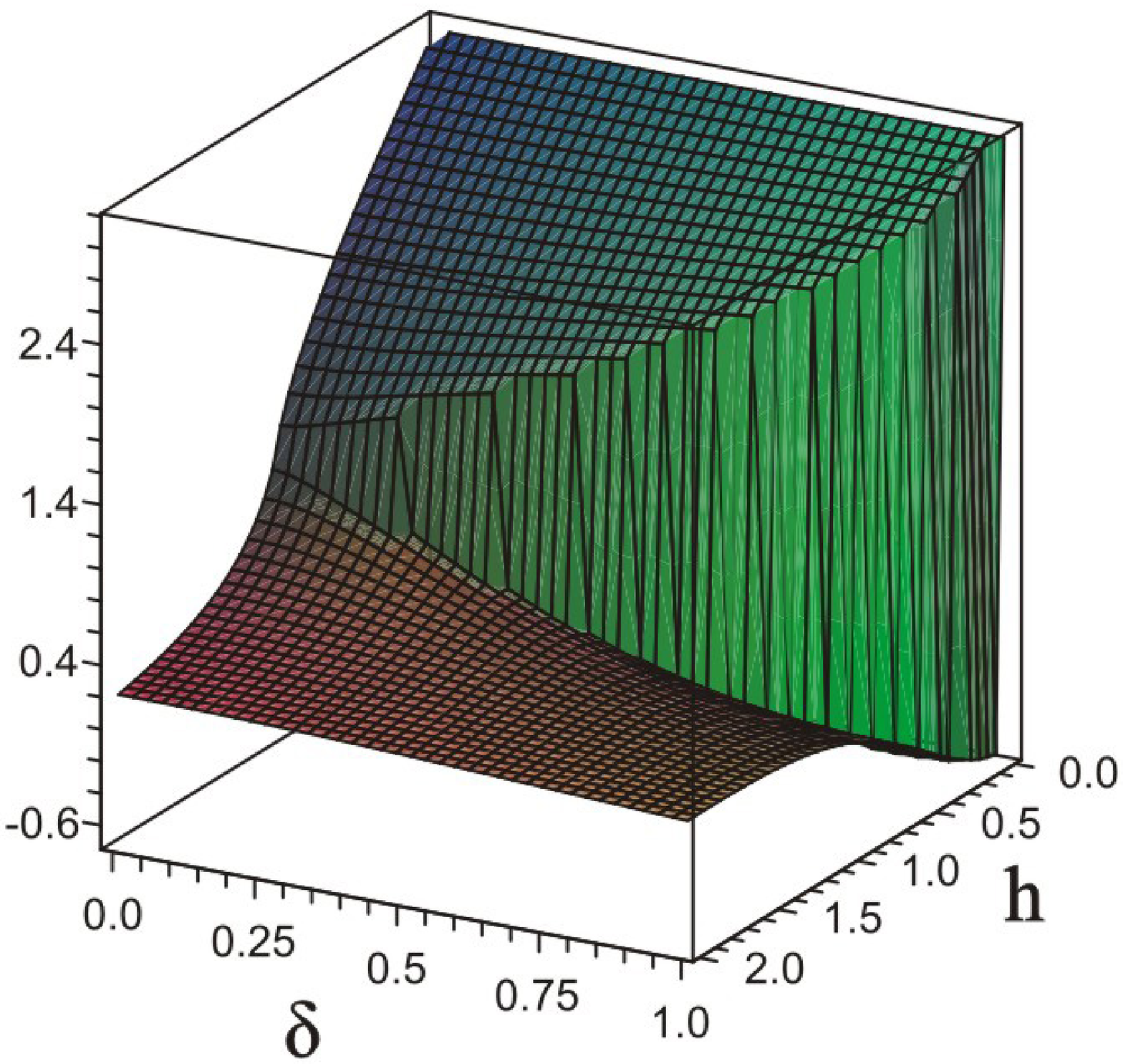}}
%\hspace{-0.75cm}
\scalebox{0.215}{\includegraphics{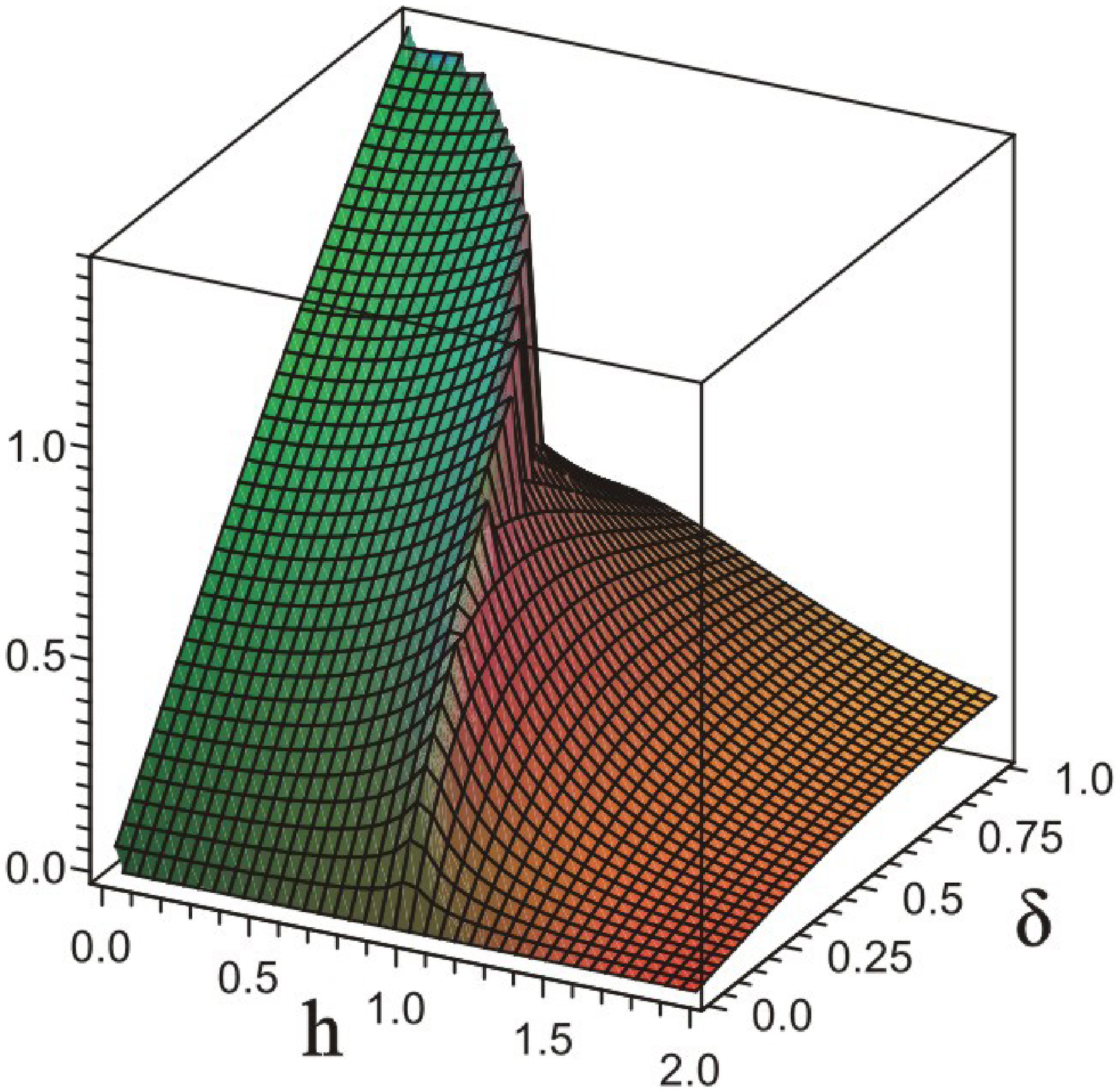}}
\end{center}
\end{minipage}
\caption{Real part of the overall geometric phase $\gamma_g$ (left) and imaginary part of the overall geometric phase $\gamma_g$ (right) versus $\delta$ and $h$.} \label{TGP1g}
\end{figure}
\begin{figure}[tbh]
\begin{minipage}[]{9 cm}
\begin{center}
\scalebox{0.225}{\includegraphics{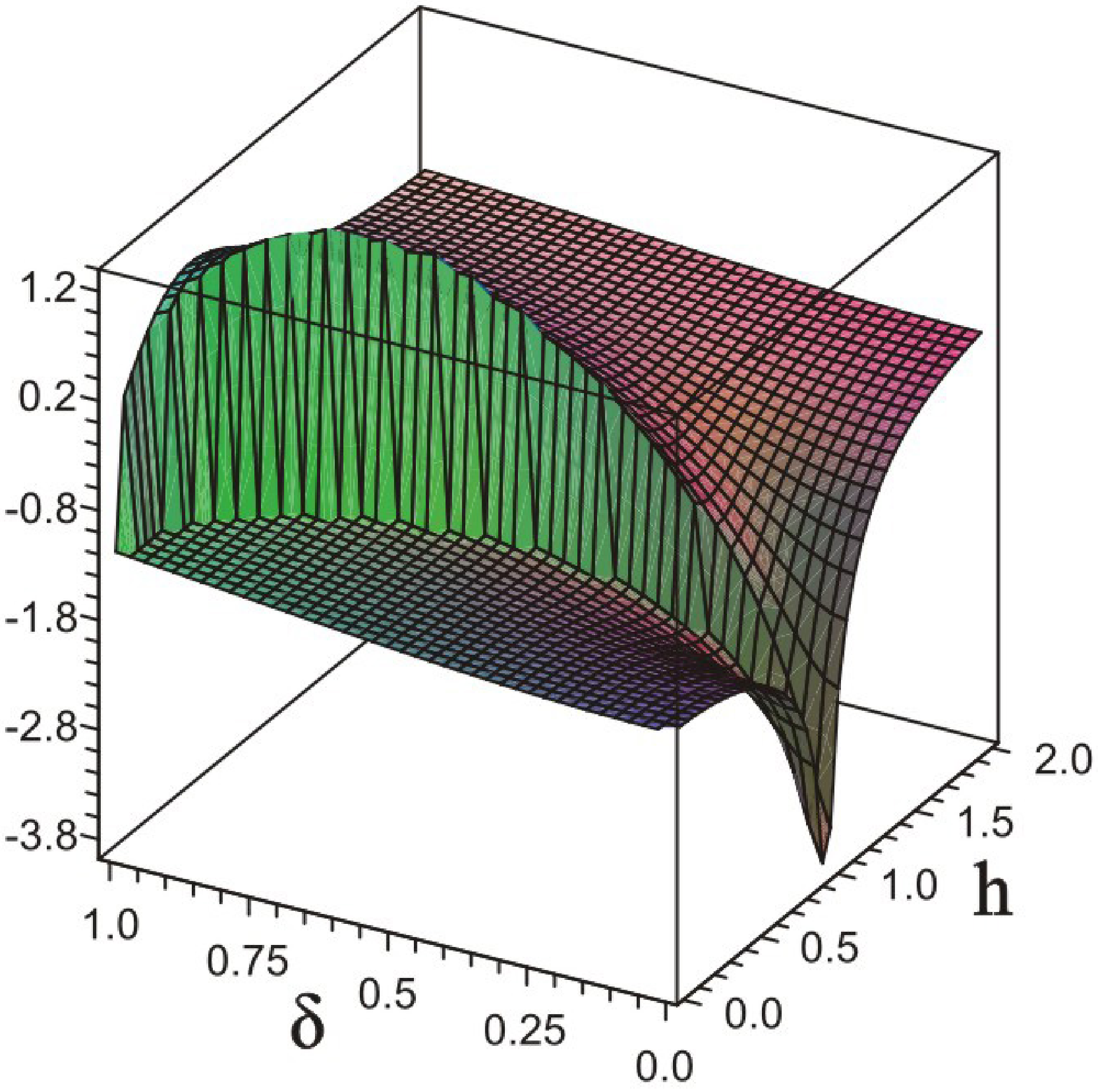}}
%\hspace{-0.75 cm}
\scalebox{0.225}{\includegraphics{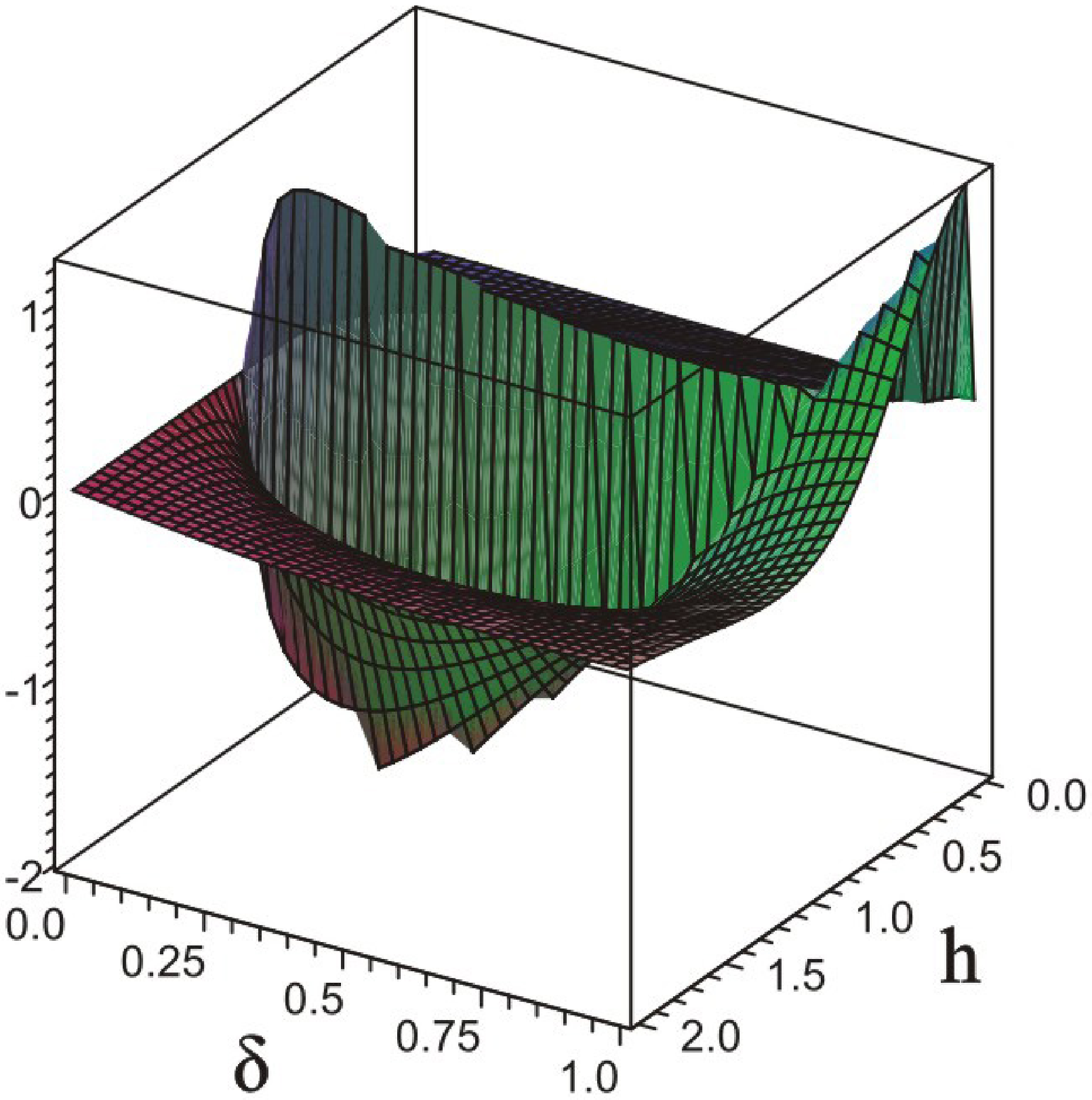}}
\end{center}
\end{minipage}
\caption{Real part of derivative of the overall geometric phase $\partial{\gamma_g}/\partial h$ (left) and its imaginary part (right) versus $\delta$ and $h$.  }
\label{DTGP1g}
\end{figure}
As can be shown, in the thermodynamical limit the energy gap $\Delta \varepsilon (h,k)$ vanishes and the geometric phase diverges at the exceptional point $h_c= (1 -\delta^2)^{1/2}$, $k_c = \arcsin \delta/a$. However, the overall geometric phase $\gamma_g= (\pi/N)\sum_{k >0}(1- \cos\theta_k)$ written in thermodynamical limit as
\begin{align}\label{B4}
\gamma_g = \int_0^{\pi}\Bigg(1 - \frac{g -  \cos x}{\sqrt{g^2 -  2g\cos x +1}} \Bigg)\, dx
\end{align}
has finite jump discontinuity at the exceptional point (Fig. \ref{TGP1g}). The result of integration can be written in terms of the complete elliptic integrals of the first and second kinds
\begin{align}\label{B5}
\gamma_g = \pi + \frac{1-g}{g} \,{\mathbf K}\bigg(\frac{2\sqrt{g}}{1+g}\bigg) -\frac{1+g}{g}\, {\mathbf E}\bigg(\frac{2\sqrt{g}}{1+g} \bigg)
\end{align}

We note that $\gamma_g$ can be written as $\gamma_g = \pi (1+ \partial E_g/\partial h)$, where $E_g = -\int_0^\pi \varepsilon(x)dx =  -iJ\delta -2(g+1)\mathbf E \big(2\sqrt{g}/(g+1)\big)$ is the ground state energy per spin. Besides, one can show that $\gamma_g = \pi(1 +  \langle \sigma^z_n\rangle)$. As known, the total magnetization per spin $\langle \sigma^z_n\rangle$ can be served as the order parameter for Ising model in a transverse magnetic field \cite{SS,PC}. This supports the statement \cite{PC,CP,Zhu,HA} that the geometric phase can be treated as the order parameter for QPT.

In Figs. \ref{TGP1g}, \ref{DTGP1g} the real and imaginary part of the overall geometric phase and its derivative as functions of external magnetic field $h$ and decay parameter $\delta$ are depicted. As can be observed $\gamma_g$ is a continuous function of $h$, if $\delta =0$, and it behaves as a step-like function, if $\delta > 0$. In the limit cases $|g| \ll 1$ and $|g| \gg 1$ we have $\rm Re \,\gamma_g \rightarrow \pi$ and $\rm Re\,\gamma_g\rightarrow 0$, respectively.

In according to the Ehrenfest classification, the QPT occurred at the exceptional point, which actually is the circle $h_c^2 + \delta_c^2=1$, is of the first order QPT. In absence of dissipation $(\delta =0)$, we have the second order QPT. Indeed, as can be observed in Figs. \ref{TGP1g}, \ref{DTGP1g}, the first derivative of the ground energy (or, equivalently, the geometric phase) is the continuous function of $h$ and its second derivative diverges at the critical point $h_c=1\, (\delta =0)$.

In summary, we establish connection between geometric phase and QPT in generic dissipative system and found the relation between the geometric phase and ground state energy. We show that the critical point where QPT occurs can be identified as the degeneracy point in the parameter space. Studying the critical behavior of the dissipative one-dimensional Ising chain in a transverse magnetic field, we find that the related QPT is of the first order QPT. In absence of dissipation it becomes the second order QPT. Our results support the claim that the relation between QPTs and geometric phase is a very general result, and the geometric phase may be considered as a good candidate to an universal order parameter for quantum phase transitions \cite{CP,Zhu}. \\

We are grateful to F. Aceves de la Cruz, A. B. Klimov, J. L. Romero and A. F. Sadreev for helpful discussions. We would also like to thank the referees for constructive comments and suggestions. This work has been supported by research grant SEP-PROMEP 103.5/04/1911 and the Programm ``Quantum macrophysics" of the Presidium of Russian Academy of Sciences.

%\bibliography{EP,nonass}
\bibliography{qpt_arxiv}

\end{document}